\documentclass[twocolumn,amsmath,prd]{revtex4}

\usepackage{epsfig}
\def\beq{\begin{equation}}
\def\eeq{\end{equation}}
\def\bea{\begin{eqnarray}}
\def\eea{\end{eqnarray}}
\def\nn{\nonumber}

\def\pslash{\not{\hbox{\kern-4pt $p$}}}
\def\qslash{\not{\hbox{\kern-4pt $q$}}}
\def\lv{\not{\hbox{\kern-4pt $L$}}}
\def\lsim{\mathrel{\raise.3ex\hbox{$<$\kern-.75em\lower1ex\hbox{$\sim$}}}}
\def\gsim{\mathrel{\raise.3ex\hbox{$>$\kern-.75em\lower1ex\hbox{$\sim$}}}}
\def\ifmath#1{\relax\ifmmode #1\else $#1$\fi}

\usepackage{xcolor}
\usepackage{hyperref}

\begin{document}

\title{Identifying Sneutrino Dark Matter: Interplay between the LHC and Direct Search}
\author{Hye-Sung Lee}
\email{hlee@bnl.gov}
\affiliation{Department of Physics, Brookhaven National Laboratory, Upton, NY 11973, USA}
\author{Yingchuan Li}
\email{ycli@quark.phy.bnl.gov}
\affiliation{Department of Physics, Brookhaven National Laboratory, Upton, NY 11973, USA}
\date{July 2011}

\begin{abstract}
Under $R$-parity, the lightest supersymmetric particle (LSP) is
stable and may serve as a good dark matter candidate. The $R$-parity
can be naturally introduced with a gauge origin at TeV scale. We go
over why a TeV scale $B-L$ gauge extension of the minimal
supersymmetric standard model (MSSM) is one of the most natural, if
not demanded, low energy supersymmetric models. In the presence of a
TeV scale Abelian gauge symmetry, the (predominantly) right-handed
sneutrino LSP can be a good dark matter candidate. Its
identification at the LHC is challenging because it does not carry
any standard model charge. We show how we can use the correlation
between the LHC experiments (dilepton resonance signals) and the
direct dark matter search experiments (such as CDMS and XENON) to
identify the right-handed sneutrino LSP dark matter in the $B-L$
extended MSSM.
\end{abstract}

\maketitle

%%%%%%%%%%%%%%%%%%%%%%%%%%%%
\section{Introduction}
\label{sec:introduction}
%%%%%%%%%%%%%%%%%%%%%%%%%%%%
There are strong evidences that about $22\%$ of the energy budget of
the Universe is in the form of dark matter (DM)
\cite{Nakamura:2010zzi}. The most precise measurement comes from
fitting the WMAP measured anisotropy of the cosmic microwave
background to the cosmological parameters \cite{Komatsu:2010fb}. One
has to rely on the other methods including direct and indirect DM
searches as well as colliders to pinpoint the identity of the DM
(see Ref. \cite{Feng:2010gw} for a review), which has far-reaching
implications for particle physics. With all standard model (SM)
particles ruled out as viable DM candidates, DM is one of the
strongest empirical evidences for the beyond SM physics.

Large Hadron Collider (LHC) at CERN will explore the physics of the
electroweak (EW) symmetry breaking and beyond. The low energy
supersymmetry (SUSY), which is one of the most popular scenarios to
stabilize the EW scale, is expected to be largely explored at the
LHC. In fact, the early search at the LHC with total energy
$\sqrt{s} = 7 ~\text{TeV}$ and integrated luminosity of $L = 35
~\text{pb}^{-1}$ has already started to put new constraints on SUSY
scenarios \cite{SUSYatLHC}.

SUSY is one of the best-motivated new physics scenarios. It can
address the gauge hierarchy problem, help unification of three SM
gauge coupling constants, and may provide a natural DM candidate.
Minimal supersymmetric standard model (MSSM) consists of the SM
fields, one more Higgs doublet and their superpartners. Typically,
the MSSM is accompanied by $R$-parity, which can protect proton from
decaying through renormalizable baryon number ($B$) or lepton number
($L$) violating terms. Under the $R$-parity, the lightest
supersymmetric particle (LSP) is stable and may serve as a DM
candidate. The MSSM provides two natural LSP DM candidates:
neutralino (superpartner of neutral gauge bosons and Higgs bosons)
and sneutrino (superpartner of neutrinos) .

The neutralino LSP DM candidate has been extensively studied and
proven to be a good DM candidate \cite{Ellis:1983ew,Jungman:1995df}.
Many studies have been done also for the detection of the neutralino
LSP signal at the collider experiments. For example, the trilepton
signals ($\chi^\pm_1 + \chi^0_2 \to 3 \ell + \text{MET}$) can be
used to look for SUSY signal with the neutralino LSP final states,
and the invariant mass distribution of dilepton ($\chi^0_2 \to
\ell^+ \ell^- + \chi^0_1$) can be used to measure superparticle
masses. (A brief summary of detecting the neutralino LSP DM signals
is included in a general SUSY review, Ref. \cite{Martin:1997ns}.)

On the other hand, the sneutrino (at earlier time, only the
left-handed one) LSP DM candidate has not been studied much, despite
of the fact it is one of only a few candidates in the SUSY scenario.
It is basically because it was excluded early as a viable DM
candidate by a combination of cosmological (DM relic density
constraint) and terrestrial constraints (direct DM search by nuclear
recoil)
\cite{Hagelin:1984wv,Falk:1994es,Ibanez:1983kw,Arina:2007tm}. The
major channel for the relic density and direct search is mediated by
the SM $Z$ boson, whose coupling to the left-handed sneutrino LSP is
too large to make it a good DM candidate.

It has been demonstrated, however, in Ref. \cite{Lee:2007mt} that
(predominantly) right-handed (RH) sneutrino ($\tilde\nu_R$) can be a
good cold DM candidate, satisfying all the constraints for viable
thermal DM candidate, when there is a TeV scale neutral gauge boson
$Z'$ that couples to the RH sneutrinos. (For an extensive review of
heavy neutral gauge boson, see Ref. \cite{Langacker:2008yv}.) There
are few studies in the RH sneutrino LSP search at the collider
experiments. Since the RH sneutrino LSP does not carry any SM
charge, we cannot use the methods developed for the neutralino LSP.
In fact, it would be very hard to see the signal related to $Z' \to
\tilde\nu_R \tilde\nu_R^*$ at the LHC experiments.

In this paper, we aim to establish a correlation between the LHC
experiments and DM direct search experiments (such as CDMS and
XENON) for a $U(1)$ gauge symmetry and discuss how we can use it to
confirm the RH sneutrino LSP DM. We choose a TeV scale $U(1)_{B-L}$
gauge symmetry. As discussed in Section \ref{sec:framework}, this is
a remarkably well-motivated (if not demanded) addition to the MSSM,
and further the economy of the model is also preserved in the sense
that we do not need the $R$-parity independently.

The rest of this paper is organized as follows. In Section
\ref{sec:framework}, we describe our theoretical framework. In
Section \ref{sec:correlation}, we discuss the correlation of the DM
direct search experiment and the LHC dilepton resonance search
experiment. In Section \ref{sec:analysis}, we show various results
of the numerical analysis. In Section \ref{sec:summary}, we
summarize our results.

%%%%%%%%%%%%%%%%%%%%%%%%%%%%
\section{Theoretical framework}
\label{sec:framework}
%%%%%%%%%%%%%%%%%%%%%%%%%%%%
Here, we describe the theoretical framework in our study.
The model we will work on is a well-known extension of the MSSM:
MSSM + three RH neutrinos/sneutrinos + TeV scale $U(1)_{B-L}$ gauge symmetry.

The RH neutrinos are well-motivated to explain the observed neutrino
masses\footnote{Supersymmetric generation of the neutrino masses,
which does not require RH neutrinos, is possible only in the absence
of the $R$-parity, which is not within our context
\cite{Hall:1983id,Grossman:1998py}.}. They are also necessary to
introduce $B-L$ as an anomaly-free gauge symmetry.

The $U(1)_{B-L}$ is one of the most popular gauge extensions as we
can see from the plethora of the literature on the subject. (For
very limited instances, see Refs.
\cite{Allahverdi:2007wt,Khalil:2007dr,Allahverdi:2008jm,Barger:2008wn,Basso:2010pe,Kajiyama:2010iq,Khalil:2011tb,Martin:1992mq}.)
It has a strong motivation especially in the SUSY framework: (i) It
is the only possible flavor-independent Abelian gauge extension of
the SM/MSSM without introducing exotic fermions (except for the RH
neutrinos which is well motivated itself by neutrino masses). (ii)
It can originate from Grand Unification Theory (GUT) models such as
$SO(10)$ and $E_6$. (iii) The radiative $B-L$ symmetry breaking,
similarly to the radiative EW symmetry breaking, in the SUSY may be
achievable \cite{Khalil:2007dr}. (iv) It can contain matter parity
$(-1)^{3(B-L)}$, which is equivalent to $R$-parity
$(-1)^{3(B-L)+2S}$, as a residual discrete symmetry
\cite{Martin:1992mq}.

In particular, the MSSM already carries the $R$-parity in order to
stabilize the proton and the LSP DM candidate. When a discrete
symmetry does not have a gauge origin, it may be vulnerable from the
Planck scale physics \cite{Krauss:1988zc}. Therefore it is more than
natural to assume a $U(1)_{B-L}$ gauge symmetry, which is a gauge
origin of the $R$-parity.

Once an Abelian gauge symmetry is introduced in the SUSY models, its
natural scale is set to be the TeV scale. This is because the masses
of sfermions (such as stop) get an extra $D$-term contribution from
a new $U(1)$ gauge symmetry and we need to make sure the sfermion
scale does not exceed the TeV scale in order to keep the SUSY as a
solution to the gauge hierarchy problem. Since much lighter scale
$U(1)$ with an ordinary size coupling should have been discovered by
the collider experiments, we can see that (roughly) TeV scale is the
right scale for the new $U(1)$ gauge symmetry in SUSY.

Therefore, replacing the $R$-parity with the TeV scale $U(1)_{B-L}$
gauge symmetry is one of the most natural and economic extensions of
the MSSM. One of the direct consequences of this model is the
existence of a TeV scale $Z'$ gauge boson, which couples to both
quarks and leptons with specific charges ($B$ for all quarks/squarks
and $-L$ for all leptons/sleptons). We assume one of the RH
sneutrinos is the LSP. It does not couple to any SM gauge boson, but
it does couple to the $Z'$ gauge boson.

It would be appropriate to comment about more general cases at this
point, before we discuss our main findings. The aforementioned
attractiveness does not exclusively apply to the $B-L$. Some mixture
with the hypercharge $Y$ (that is, $(B-L) + \alpha Y$ with some
constant $\alpha$) or lepton flavor dependent $U(1)$ gauge symmetry
($B-x_iL$) \cite{Lee:2010hf} are also known to be anomaly-free
without introducing exotic fermions, and can have the matter parity
as a residual discrete symmetry. (For some references about discrete
symmetries from a gauge origin, see Refs.
\cite{Ibanez:1991hv,Ibanez:1991pr,Dreiner:2005rd,Hur:2008sy,Lee:2008zzl}.)
It would not be difficult to distinguish them with the LHC
experiments though. The forward-backward asymmetry can tell about
the $Z'$ couplings \cite{Langacker:1984dc,Barger:1986hd}. The $B-L$
is vectorial which can distinguish itself from the axial coupling
provided by the $Y$ in the forward-backward asymmetry measurement.
The lepton flavor dependence of couplings can be easily seen by
comparing the dilepton $Z'$ resonance signals \cite{Lee:2010hf}.

%%%%%%%%%%%%%%%%%%%%%%%%%%%%
\section{Correlation of two experiments}
\label{sec:correlation}
%%%%%%%%%%%%%%%%%%%%%%%%%%%%
%%% Figure %%%
\begin{figure}[tb]
\begin{center}
\includegraphics[height=3.0cm]{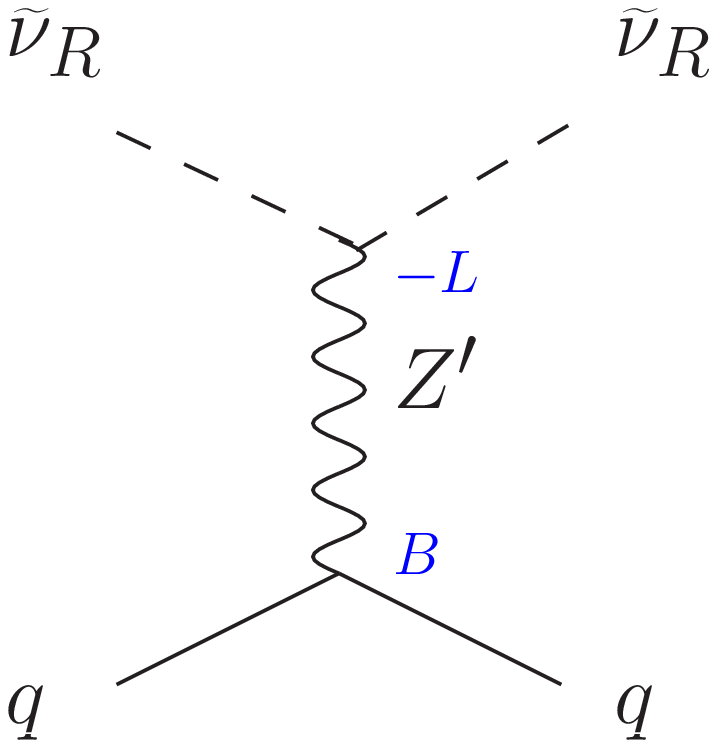}
~~
\includegraphics[height=2.35cm]{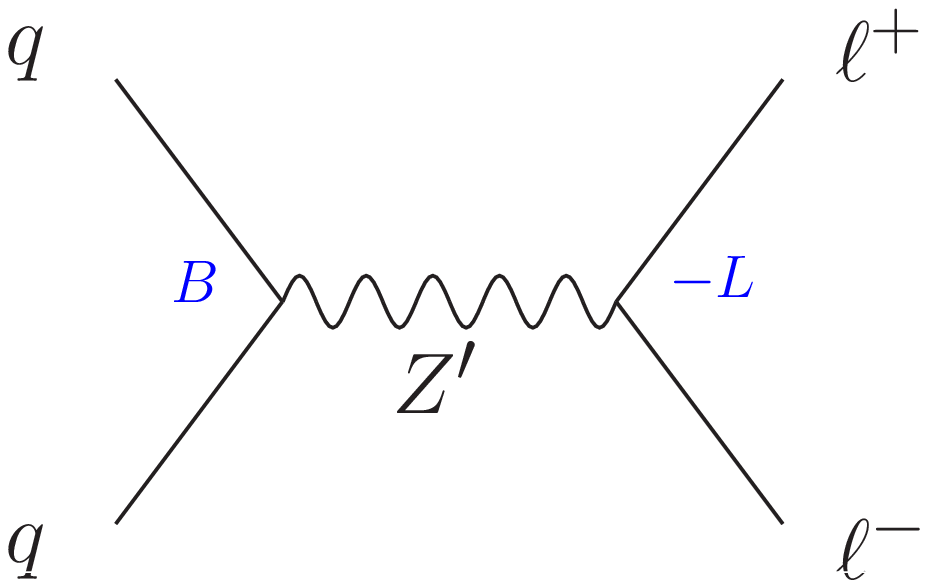} \\
(a) ~~~~~~~~~~~~~~~~~~~~~~~~~~~~~~ (b) ~~~~~~
\end{center}
\caption{\label{fig:diagrams}
(a) Sneutrino LSP dark matter direct search using nuclear recoil. (b) Dilepton $Z'$ resonance at the LHC.}
\end{figure}
%%%%%%%%%%%%%%

In this section, we discuss the interplay between two experiments:
the dilepton $Z'$ search at the LHC and the direct DM search
experiments.

We will not consider the relic density constraints in our study. We
are mainly interested in establishing the correlation between the
LHC and the direct DM search with minimal assumptions. The relic
density constraint in principle depends on the cosmological
assumptions (for example, whether the DM was thermally in
equilibrium in the early Universe or not). Furthermore, the channels
to reproduce the right DM relic density are not unique: it may
involve $Z'$ as well as its superpartner $\tilde Z'$. The former
suggests the RH sneutrino LSP DM mass is quite close to a half of
$Z'$ mass, but the latter does not suggest it. (See Ref.
\cite{Lee:2007mt} for details.) However, once the RH sneutrino is
confirmed by our suggested interplay of the LHC and the direct DM
search, one can compare the measured DM mass with those that can
satisfy the relic density constraint to test consistency with the
standard cosmology.

The direct DM search experiments such as CDMS \cite{Ahmed:2009zw}
and XENON \cite{Aprile:2011hi} can detect the DM by observing the
signal from the nuclear recoil. For the RH sneutrino LSP DM, which
is a SM singlet, it is mediated by the $Z'$. (See Fig.
\ref{fig:diagrams} (a).) Following the approach of Ref.
\cite{Lee:2007mt}, we can see that the effective Lagrangian for the
direct DM search in our framework is given by \beq {\cal L} = i
\frac{g^2_{Z'}}{M^2_{Z'}} \left(-1\right) \left( {\tilde \nu}^*_R
\partial_{\mu}{\tilde \nu}_R-\partial_{\mu}{\tilde \nu}^*_R {\tilde
\nu}_R \right) \sum_{i=u,d} \left(\frac{1}{3}\right) {\bar q}_i
\gamma_{\mu} q_i \eeq

The spin-independent cross section per nucleon via a $Z'$ gauge
boson exchange, in the non-relativistic limit, is given by \beq
\sigma^\text{SI}_\text{nucleon} = \frac{\left(Z \lambda_p + (A-Z)
\lambda_n\right)^2}{\pi A^2} \mu_n^2 \label{eq:sigmaDirect} \eeq
where the $\mu_n$ ($\simeq m_\text{proton}$ for $m_{\tilde \nu_R}
\gg m_\text{proton}$) is the effective mass of the nucleon and the
DM. In general, the $u$ and $d$ quarks would have different
couplings to the $Z'$, and the cross section would depend on the
detector type. Under $B-L$, however, the $u$ and $d$ quarks carry
the same charge, and the $Z'$ coupling to proton and neutron are the
same $\lambda_p = \lambda_n = -\frac{g_{Z'}^2}{M_{Z'}^2}$. Thus Eq.
\eqref{eq:sigmaDirect} has a simple form of \beq
\sigma^\text{SI}_\text{nucleon} =
\left(\frac{g_{Z'}^2}{M_{Z'}^2}\right)^2 \frac{\mu_n^2}{\pi} \eeq
which depends only on the $g_{Z'}/M_{Z'}$ regardless of the detector
type.

The process at LHC that is directly correlated with the direct
search is the di-sneutrino $Z'$ resonance process ($q \bar q \to Z'
\to \tilde\nu_R \tilde\nu_R^*$), whose observation would be
practically impossible since it does not leave anything but the
missing energy. Nevertheless, a typical dilepton $Z'$ resonance ($q
\bar q \to Z' \to \ell^+ \ell^-$) can reveal the relevant
information, because all leptons and sleptons carry the same charge
($-L$), though the spin and mass of the final particles are
different. (See Fig. \ref{fig:diagrams} (b).) If we neglect the
effect of the analysis cuts, the dilepton $Z'$ resonance cross
section for the $B-L$ model is determined by 3 parameters: mass of
$Z'$ ($M_{Z'}$), width of $Z'$ ($\Gamma_{Z'}$), and gauge coupling
constant ($g_{Z'}$).

The details of the dilepton $Z'$ resonance at the hadron collider
was elegantly analyzed in Ref. \cite{Carena:2004xs} although the
focus was given for the $p \bar p$ collider. In the narrow width
approximation, one can write down the dilepton $Z'$ resonance cross
section as \bea
&& \sigma_\text{Dilepton} \nn \\
&\equiv& \sigma(p p \to Z' \to \ell^+ \ell^- ) \\
&=& \frac{\pi g_{Z'}^2}{48 s} \left[ 2 \cdot
\left(\frac{1}{3}\right)^2 w_u + 2 \cdot \left(\frac{1}{3}\right)^2
w_d \right] \text{Br} (Z' \to \ell^+ \ell^-) \nn \eea where the
functions $w_u$ and $w_d$ includes\ the parton distribution function
information for the $u$ and $d$ quarks, respectively. (See Ref.
\cite{Carena:2004xs} for details.) The branching ratio can be
written as \beq \text{Br} (Z' \to \ell^+ \ell^-) = \frac{g_{Z'}^2
M_{Z'}}{24 \pi \Gamma_{Z'}} \left[ 2 \cdot (-1)^2 \right]. \eeq

With $M_{Z'}$ and $\Gamma_{Z'}$ fixed, the $\sigma_{{\rm Dilepton}}$
is proportional to $g^4_{Z'}$, the same dependence as the direct
detection cross section $\sigma^{{\rm SI}}_{{\rm nucleon}}$. While
$\sigma^{{\rm SI}}_{{\rm nucleon}}$ is proportional to
$M^{-4}_{Z'}$, the $\sigma_{{\rm Dilepton}}$ carries different and
more complicated dependence on the mass $M_{Z'}$. The contribution
to the $\sigma_{{\rm Dilepton}}$ from the $Z'$ propagator is
$[(M^2_{l^+l^-}-M^2_{Z'})^2+M^2_{Z'}\Gamma^2_{Z'}]^{-1} \approx \pi
\delta(M^2_{l^+l^-}-M^2_{Z'})/M_{Z'} \Gamma_{Z'}$ in the narrow
width approximation. The dependence of $\sigma_{{\rm Dilepton}}$ on
parton distribution functions further makes the $M_{Z'}$ dependence
more complicated. Moreover, the $\sigma_{{\rm Dilepton}}$ also
depends on the total width $\Gamma_{Z'}$, which is an irrelevant
parameter for $\sigma^{{\rm SI}}_{{\rm nucleon}}$.

An appropriate quantity for the examination of the correlation is
the ratio of two cross sections $\sigma^{{\rm SI}}_{{\rm
nucleon}}/\sigma_{{\rm Dilepton}}$. The gauge coupling cancels and
the ratio only depends on the mass and width of $Z'$. In practice,
with signal events observed, the mass and total width can be
determined by fitting the resonance peak to the Breit-Wigner form
$1/[(M^2_{l^+l^-}-M^2_{Z'})^2+M^2_{Z'}\Gamma^2_{Z'}]$. Thus, we can
confirm the RH sneutrino LSP DM by checking if the experimental
results and theoretical predictions of the $\sigma^{{\rm SI}}_{{\rm
nucleon}}/\sigma_{{\rm Dilepton}}$ are consistent. (We will discuss
it further in the following section.) This method to identify the RH
sneutrino LSP DM using the interplay of the LHC and the direct DM
search experiments is our main finding in this paper.

Before the presentation of numerical analysis in the next section,
we briefly comment about the experimental bounds and the LHC
discovery potential of the model here. A dedicated study of this has
been carried out in Ref. \cite{Basso:2010pe}, where the bounds on
$g_{Z'}$ and $M_{Z'}$ from LEP \cite{Cacciapaglia:2006pk} and recent
Tevatron search \cite{Abazov:2010ti,Aaltonen:2011gp} have been
discussed \footnote{We notice the recent searches
\cite{Aad:2011xp,Chatrchyan:2011wq} at the LHC at 7 TeV with
integrated luminosity of 40 pb$^{-1}$ put slightly stronger bounds
than the Tevatron search.}, and the reaches at LHC of 7, 10, and 14
TeV with various luminosity have been explored. According to Ref.
\cite{Basso:2010pe}, the LHC will probe a large portion of the
region with $g_{Z'}$ larger than 0.01 and $M_{Z'}$ within a few TeV.
The value of $\sigma^{{\rm SI}}_{{\rm nucleon}}$ in the major
portion of such parameter region is larger than $10^{-48} [{\rm
cm}^2]$. It would be explored by the upcoming direct detection
experiments, at SNOLAB and DUSEL for instance, if their precision
can be improved by another 2 to 3 orders of magnitude beyond the
current most stringent bounds from XENON100 \cite{Aprile:2011hi}. We
therefore conclude that there is a large common region in the
$g_{Z'}-M_{Z'}$ plane that will be probed at both experiments. It is
thus possible to test the model by the correlations of these two
phenomenological aspects.

%%%%%%%%%%%%%%%%%%%%%%%%%%%%
\section{Numerical Analysis}
\label{sec:analysis}
%%%%%%%%%%%%%%%%%%%%%%%%%%%%

\begin{figure}
\centerline{
\put(-125,-55){\epsfig{file=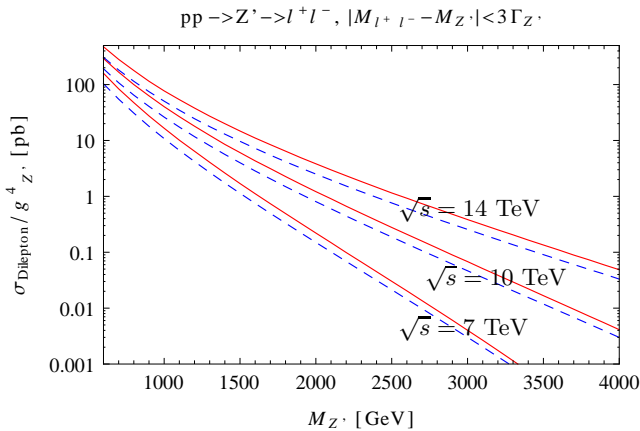,width=8.5cm,angle=0} }
\put(25,25){$\sqrt{s}=14$ TeV} \put(35,-1){$\sqrt{s}=10$ TeV}
\put(25,-20){$\sqrt{s}=7$ TeV} } \caption{\label{fig:LHCxsec} The
coupling normalized cross section of $p p \rightarrow Z' \rightarrow
l^+l^-$ ($l=e,\mu$) with invariant mass cut
$|M_{l^+l^-}-M_{Z'}|<3\Gamma_{Z'}$ imposed in $U(1)_{B-L}$ model at
the LHC with $\sqrt{s}=7$, 10, and 14 TeV. The $Z'$ width
$\Gamma_{Z'}$ is taken as 3$\%$ (red solid line) and 6$\%$ (blue
dashed line) of the mass $M_{Z'}$. }
\end{figure}

In the following, we discuss the dilepton resonance production cross
section $\sigma_{{\rm Dilepton}}$ and the ratio $\sigma^{{\rm
SI}}_{{\rm nucleon}}/\sigma_{{\rm Dilepton}}$ as functions of the
mass $M_{Z'}$ for different values of width $\Gamma_{Z'}$.

Taking into account the decay modes to SM particles only, we find
the width of $Z'$ is roughly $\Gamma^{{\rm SM}}_{Z'}\approx
0.2g^2_{Z'}M_{Z'}$. With all the possible decay channels included,
the total width $\Gamma_{Z'}$ depends on the full mass spectrum,
with the $\Gamma^{{\rm SM}}_{Z'}$ setting the minimum value. For
illustration purpose, we will take $\Gamma_{Z'}/M_{Z'}=3\%$ and
6$\%$ in the analysis.

For the simulation of the dilepton resonance production process $pp
\rightarrow Z' \rightarrow l^+ l^-$ at the LHC, we use the CTEQ6.1L
parton distribution functions \cite{Pumplin:2002vw}. We adopt the
event selection criteria with the basic cuts \cite{LHCcuts}
\begin{equation}
\label{eq:cuts1} p_{T_{l}} > 20 ~{\rm GeV}, ~~|\eta_{l}| < 2.5,
\end{equation}
and we further impose cut on the invariant mass of lepton pair
\begin{equation}
\label{eq:cuts2} |M_{l^+ l^-}-M_{Z'}|<3 \Gamma_{Z'}.
\end{equation}
The cross sections $\sigma_{{\rm Dilepton}}$ normalized by gauge
coupling for the process $pp \rightarrow Z' \rightarrow l^+ l^-$ at
the LHC of 7, 10, and 14 TeV, with cuts in Eq. (\ref{eq:cuts1}),
(\ref{eq:cuts2}) imposed, are shown in Fig. \ref{fig:LHCxsec}.

\begin{figure}
\centerline{
\put(-125,-55){\epsfig{file=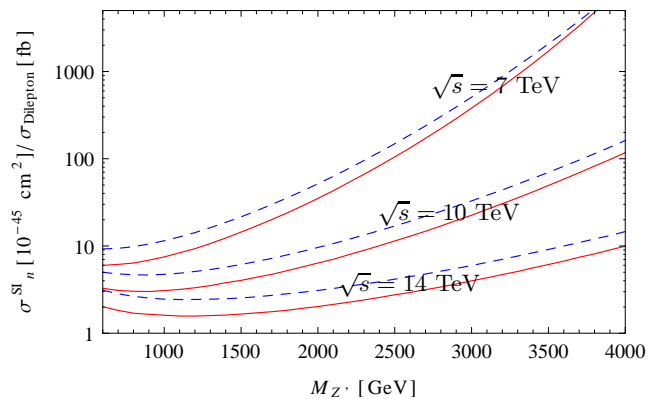,width=8.5cm,angle=0} }
\put(35,60){$\sqrt{s}=7$ TeV} \put(15,12){$\sqrt{s}=10$ TeV}
\put(0,-15){$\sqrt{s}=14$ TeV} } \caption{\label{fig:LHCDMratio} The
ratio of cross sections of the spin-independent sneutrino-nucleus
elastic scattering ${\tilde \nu}_R q \rightarrow {\tilde \nu}_R q$
(normalized to a single nucleon) at the DM direct detection
experiments and the process $pp \rightarrow Z' \rightarrow l^+
l^-(|M_{l^+ l^-}-M_{Z'}|<3 \Gamma_{Z'})$ at the LHC at 7, 10, and 14
TeV. The $Z'$ width $\Gamma_{Z'}$ is taken as 3$\%$ (red solid line)
and 6$\%$ (blue dashed line) of the mass $M_{Z'}$.}
\end{figure}

In Fig. \ref{fig:LHCDMratio}, we show the ratios $\sigma^{{\rm
SI}}_{{\rm nucleon}}/\sigma_{{\rm Dilepton}}$, for various center of
mass energy 7, 10, and 14 TeV at the LHC, as functions of $M_{Z'}$
for $\Gamma_{Z'}/M_{Z'}=3\%$, $6\%$. As the gauge coupling cancels,
the ratio only depends on the mass $M_{Z'}$ and width $\Gamma_{Z'}$
of $Z'$.

The future direct detection experiments will reach the sensitivity
beyond $10^{-45}$ cm$^2$ level. The future running of LHC at 7, 10,
and 14 TeV will have integrated luminosity ranging from a few
fb$^{-1}$ to a few 100 fb$^{-1}$. Assuming the background is
negligible compared to the signal as is the case here, the discovery
at LHC at 3$\sigma$ and 5$\sigma$ significance requires 5 and 15
events, respectively. The LHC with integrated luminosity of 1
fb$^{-1}$ (100 fb$^{-1}$) will be able to probe the cross section at
10 fb (0.1 fb) level. If positive signals are observed at both
experiments, and they obey the predicted ratio as shown in Fig.
\ref{fig:LHCDMratio}, it should be taken as a rather strong hint for
the sneutrino LSP DM scenario. Otherwise, the model can be ruled out
if positive signals are observed in either or both experiments but
not consistent with the predicted ratio shown in Fig.
\ref{fig:LHCDMratio}.

The mass and width of $Z'$ need to be determined from the LHC data
for the purpose of this examination of the ratio of cross sections.
Since the momentum resolution of $e^{\pm}$ is better than
$\mu^{\pm}$ in the high $P_T$ region, the $e^+e^-$ final state is
more favorable than the $\mu^+\mu^-$ final state for this purpose.
There are errors in the determination of width arising from momentum
resolution as well as fitting to the Breit-Wigner form with limited
number of events. A quantitative study on these errors is beyond the
scope of this paper. However, these need to be considered when a
comparison of cross sections is carried out in the future after
positive signals are observed.

%%%%%%%%%%%%%%%%%%%%%%%%%%%%
\section{Summary}
\label{sec:summary}
%%%%%%%%%%%%%%%%%%%%%%%%%%%%

We study the sneutrino LSP DM scenario in the SUSY $U(1)_{B-L}$
model at the LHC and direct detection experiments. The sneutrino
only couples to the $Z'$, making it extremely hard to test this
model at the LHC. However, since charged leptons and sneutrinos
carry the same $B-L$ charge, the charged lepton $e^{\pm}, \mu^{\pm}$
can serve as a good replacement of sneutrino for diagnosing purpose.

Following this spirit, we propose to test this scenario at the LHC
with the process $pp \rightarrow Z' \rightarrow l^+ l^-(l=e,\mu)$.
The cross section of this process is tightly correlated with that of
the sneutrino-nucleus spin-independent elastic scattering in the
direct detection experiments. Since a large common region of the
parameter space will be probed by both experiments, the correlation
can be used to confirm or rule out such model. In particular, with
the signal events of dilepton resonance production observed at the
LHC and with the $Z'$ mass and width extracted from the data, the
ratio $\sigma^{{\rm SI}}_{{\rm nucleon}}/\sigma_{{\rm Dilepton}}$ is
fixed in this scenario and can be examined against the experimental
data.

%%%%%%%%%%%%%%%%%%%%%%%%%%%%
\acknowledgments
%%%%%%%%%%%%%%%%%%%%%%%%%%%%
{\it Acknowledgments:} We thank T. Han and F. Paige for helpful
discussions. We further thank T. Han for providing the Fortran code
HANLIB that is used in the Monte Carlo simulations. This work is
supported by the US DOE  under Grant Contract DEAC02-98CH10886.

%%%%%%%%%%%%%%%%%%%%%%%%%%%%

\end{document}